\documentclass[sigconf]{acmart}

\usepackage{booktabs} 
\usepackage{stfloats} 

\definecolor{shadecolor}{gray}{0.9}
\definecolor{Green}{HTML}{156946}
\definecolor{Strawberry}{HTML}{D13F8A}

\hypersetup{citecolor=Green}
\hypersetup{linkcolor=Green}
\hypersetup{urlcolor=Green}

\usepackage{dsfont}
\usepackage{nicefrac}

\copyrightyear{2018} 
\acmYear{2018} 
\setcopyright{rightsretained} 
\acmConference[RecSys '18]{Twelfth ACM Conference on Recommender Systems}{October 2--7, 2018}{Vancouver, BC, Canada}
\acmBooktitle{Twelfth ACM Conference on Recommender Systems (RecSys '18), October 2--7, 2018, Vancouver, BC, Canada}
\acmDOI{10.1145/3240323.3240370}
\acmISBN{978-1-4503-5901-6/18/10}

\makeatletter
\g@addto@macro\normalsize{%
  \setlength\abovedisplayskip{4pt}
  \setlength\belowdisplayskip{4pt}
  \setlength\abovedisplayshortskip{2pt}
  \setlength\belowdisplayshortskip{2pt}
}
\makeatother

\begin{document}
\title[Algorithmic Confounding in Recommendation Systems Increases Homogeneity]{How Algorithmic Confounding in Recommendation Systems Increases Homogeneity and Decreases Utility}

\author{Allison J.B.~Chaney}
\affiliation{%
 \institution{Princeton University}
 \department{Department of Computer Science}
}
\email{achaney@princeton.edu}

\author{Brandon M. Stewart}
\affiliation{%
 \institution{Princeton University}
 \department{Department of Sociology}
}
\email{bms4@princeton.edu}

\author{Barbara E. Engelhardt}
\affiliation{%
 \institution{Princeton University}
 \department{Department of Computer Science}
}
\email{bee@princeton.edu}

\begin{abstract}
Recommendation systems are ubiquitous and impact many domains; they have the potential to influence product consumption, individuals' perceptions of the world, and life-altering decisions.  These systems are often evaluated or trained with data from users already exposed to algorithmic recommendations; this creates a pernicious feedback loop.
Using simulations, we demonstrate how using data confounded in this way homogenizes user behavior without increasing utility.\looseness=-1
\end{abstract}

\settopmatter{printacmref=false}

\maketitle

\section{Introduction}
\begin{figure}[t]
\includegraphics[width=0.9\columnwidth]{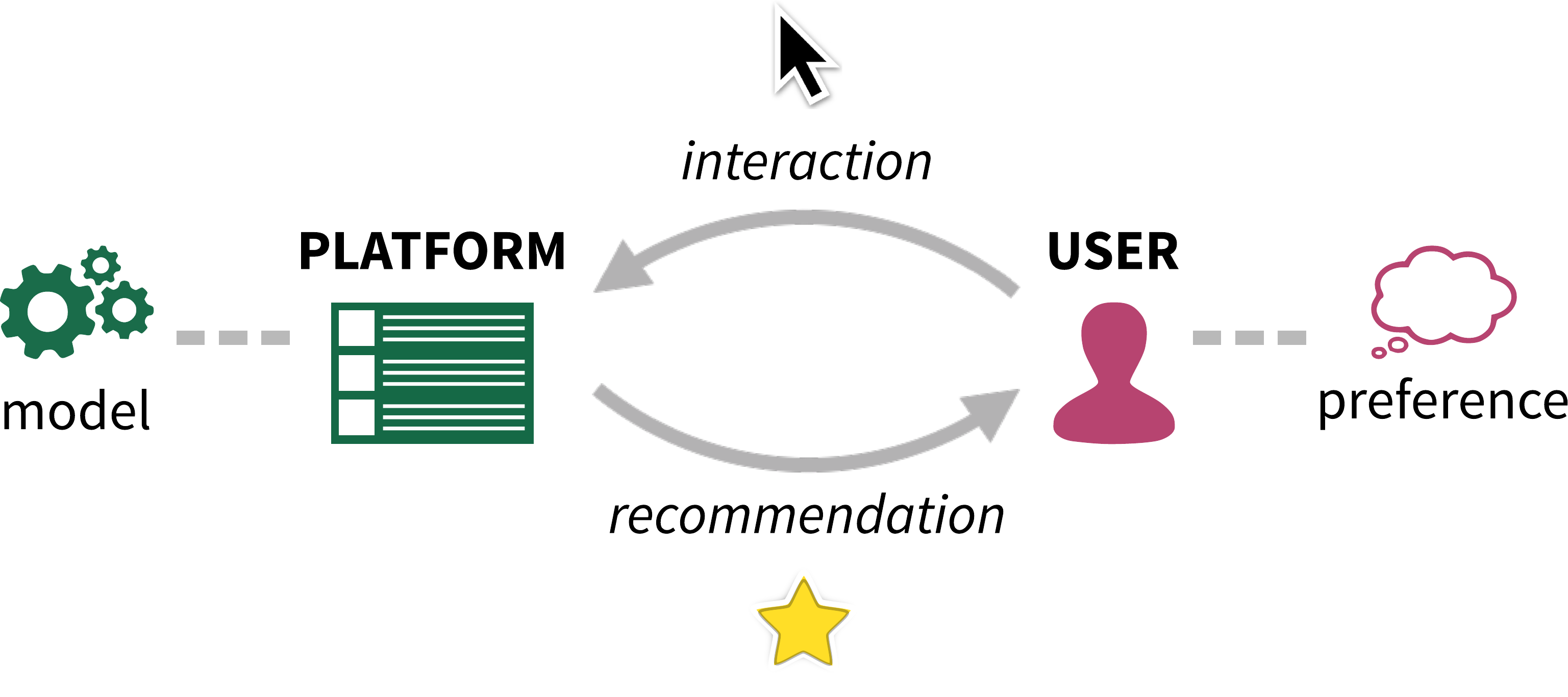}
\caption{The feedback loop between user behavior and algorithmic recommendation systems.  Confounding occurs when a platform attempts to model user behavior without accounting for recommendations.  User preferences act as confounding factors, influencing both recommendations (through past interactions) and current interactions.  \looseness=-1}
\label{fig:cartoon}
\end{figure}

Recommendation systems are designed to help people make decisions.  These systems are commonly used on online platforms for video, music, and product purchases through service providers such as Netflix, Pandora, and Amazon.  Live systems are updated or retrained regularly to incorporate new data that was influenced by the recommendation system itself, forming a feedback loop (\cref{fig:cartoon}).
While the broad notion of confounding from the data collection process has been studied extensively, we seek to \emph{characterize} the impact of this feedback loop in the context of recommendation systems, demonstrating the unintended consequences of \emph{algorithmic confounding}.
As recommendation systems become increasingly important in decision-making, we have an ethical responsibility to understand the idiosyncrasies of these systems and consider their implications for individual and societal welfare~\cite{2018BetterRecs}.\looseness=-1

Individual decisions can aggregate to have broad political and economic consequences. Recommendation systems can influence how users perceive the world by filtering access to media; pushing political dialog towards extremes~\cite{2018YouTubeRadicalizer} or filtering out contrary opinions~\cite{2017NYT}.
Even more gravely, these systems impact crucial decision-making processes, such as loan approvals, criminal profiling, and medical interventions.  As recommendation systems shape access to goods and resources, issues of fairness and transparency need to be considered.  For example, if a distinct group of minority-preference users exists, a system may deferentially undermine the needs or preferences of the minority in favor of optimizing the utility of the majority group.\looseness=-1

Many researchers and practitioners still focus on evaluating recommendation systems in terms of held-out accuracy, which cannot capture the full effects of the feedback loop.  Even with accuracy as a primary concern, algorithmic confounding can play a crucial role; for instance, when a recommendation system is evaluated using confounded held-out data, results are biased toward recommendation systems similar to the confounding algorithm.  Thus, the choice of data can considerably impact held-out evaluation and subsequent conclusions.  Averaged accuracy metrics, however, are only one approach to evaluating recommendation systems, and do not detect disparate impact across users.  It is our hope that this work will help motivate researchers and practitioners to 1) actively assess systems with objectives such as diversity, serendipity, novelty, and coverage~\cite{kaminskas2017diversity}; 2) apply causal reasoning techniques to counter the effects of algorithmic confounding; and 3) evaluate the distribution of impact across all users, instead of exclusively reporting averages.\looseness=-1

We begin with a summary of our claims (\cref{sec:claims}) and then situate this work among related lines of inquiry (\cref{sec:related}).
To provide evidence for our claims, we introduce a model for users interacting with recommendations (\cref{sec:interaction}); this allows us to analyze the impact of algorithmic confounding on simulated communities (\cref{sec:simulations}).
We find that algorithmic confounding amplifies the homogenization of user behavior (\cref{sec:homogenization}) without corresponding gains in utility (\cref{sec:utility}) and also amplifies the impact of recommendation systems on the distribution of item consumption (\cref{sec:gini}).
We briefly discuss weighting approaches to counter these effects (\cref{sec:confounded_eval}) before we conclude (\cref{sec:conclusion}).  For our simulations and analysis, we develop a general framework for recommendation systems (\cref{sec:framework}) which highlights commonalities between seemingly distinct approaches. 
\looseness=-1

\section{Consequences of the Feedback Loop}
\label{sec:claims}
Real-world recommendation systems are often part of a feedback loop (\cref{fig:cartoon}): the underlying recommendation model is trained using data that are confounded by algorithmic recommendations from a previously deployed system.  
We attempt to \emph{characterize} the impact of this feedback loop through three claims.

\vspace{5pt}
\textit{The recommendation feedback loop causes homogenization of user behavior} (\cref{sec:homogenization}), which is amplified with more cycles through the loop.  Homogenization occurs at both a population level (all users behave more similarly) and at an individual level (each user behaves more like its nearest neighbors).

\vspace{5pt}
\textit{Users experience losses in utility due to homogenization effects; these losses are distributed unequally} (\cref{sec:utility}).  Across multiple recommendation system types, users with lower relative utility have higher homogenization, indicating that homogenization can be detrimental for subsets of users by encouraging them to make sub-optimal choices.\looseness=-1

\vspace{5pt}
\textit{The feedback loop amplifies the impact of recommendation systems on the distribution of item consumption} (\cref{sec:gini}), irrespective of homogenization effects. Two recommendation systems can produce similar amounts of user homogenization with different distributions of item consumption.\looseness=-1

\section{Related Work}
\label{sec:related}

\paragraph{Bias, confounding, and estimands}
\label{sec:estimands}
Schnabel, et al.~\cite{schnabel2016recommendations}
note that users introduce selection bias; this occurs during the interaction component of the feedback loop shown in \cref{fig:cartoon}. They consider a mechanism for interaction in which users first select an item and then rate it.  Other work also considers similar notions of
missingness in rating data \cite{marlin2009collaborative, ying2006leveraging}.
However, many platforms exist where users express their preferences implicitly by viewing or reading content, as opposed to explicitly rating it.
In the implicit setting, the observations of user behavior are the selections themselves.  The quantity of interest (estimand) is no longer the rating of an item, but the probability of the user selecting an item.  Thus, we do not wish to correct for user selection bias; instead, we wish to predict it.\looseness=-1

Recommendation systems introduce confounding factors in this setting; it is difficult to tell which user interactions stem from users' true preferences and which are influenced by recommendations.  The core problem is that recommendation algorithms are attempting to model the underlying user preferences, making it difficult to make claims about user behavior (or use the behavior data) without accounting for algorithmic confounding.  In this paper, we describe various problems that arise from using data confounded by recommendation systems.
Among these problems is offline evaluation using confounded data; De Myttenaere, et al.~\cite{de2014reducing}
propose addressing this with weighting techniques and Li, et al.~\cite{li2011unbiased} propose a method specifically for reducing this bias in contextual bandit algorithms.\looseness=-1

Previous work has investigated how many algorithms rely on data that is imbued with societal biases and explored how to address this issue~\cite{baeza2016data,Sweeney:2013,chander2016racist,sandvig2014auditing}.  This work is complementary to these efforts as the described feedback effects may amplify societal biases.
Due to these and other concerns, regulations are emerging to restrict automated individual decision-making, such as recommendation systems~\cite{goodman2016european}; this work will aid in making such efforts effective.

\paragraph{Evaluating recommendation systems} 
Rating prediction is a common focus for recommendation algorithms, owing its popularity at least in part to the Netflix challenge \cite{bennett2007netflix,bell2007lessons}, which evaluated systems using RMSE on a set of held-out data.  In practice, however, recommendation systems are deployed to rank items.  Even Netflix has moved away from its original 5-star system in favor of a ranking-friendly thumbs-up/down interaction~\cite{youtubeThumbs} and now advocates for ranking items as the primary recommendation task, as it considers \emph{all items in a collection} during evaluation instead of \emph{only held-out observed items} \cite{steck2013evaluation}.
Simply put, top-$n$ correctness metrics such as precision, recall, and nDCG are better for evaluating the real-world performance of these systems \cite{cremonesi2010performance}.
These metrics are popular because they are straightforward to understand and easy to implement, but still they do not necessarily capture the usefulness of recommendations; a variety of system characteristics are thought to be important~\cite{herlocker2004evaluating}.
Among these characteristics, diversity is often framed as a counterpart to accuracy \cite{zhou2010solving,javari2015probabilistic,liu2012solving}.
Diversity can be considered at multiple levels: in aggregate, within groups of users, and individually.  Many efforts have been made to understand whether or not various recommender systems impact diversity by reinforcing the popularity of already-popular products or by increasing interactions with niche items~\cite{fleder2009blockbuster,anderson2006long,celma2008hits,park2008long,zeng2015modeling,dan2013long}.
These systems also have the potential to create ``echo-chambers,'' which result in polarized behavior \cite{dandekar2013biased}.\looseness=-1

\paragraph{Causality in recommendation systems}
Formal causal inference techniques have been used extensively in many domains, but have only recently been applied to recommendation systems.
Liang, et al.~\cite{liang2016modeling} draw on the language of causal analysis in describing a model of user exposure to items; this is related to distinguishing between user preference and our confidence in an observation~\cite{Hu08collaborativefiltering}.
Some work has also been done to understand the causal impact of these systems on behavior by finding natural experiments in observational data \cite{Sharma:2015,su2016effect} (approximating expensive controlled experiments~\cite{kohavi2009controlled}), but it is unclear how well these results generalize.
Schnabel, et al.~\cite{schnabel2016recommendations} use propensity weighting techniques to remove users' selection bias for explicit ratings.
Bottou, et al.~\cite{bottou2013counterfactual} use ad placement as an example to motivate the use of causal inference techniques in the design of deployed learning systems to avoid confounding; this potentially seminal work does not, however, address the use of already confounded data (e.g., to train and evaluate systems or ask questions about user behavior), which is abundant.\looseness=-1

\paragraph{Connections with the explore/exploit trade-off}
Some investigators have modeled these systems using temporal dynamics \cite{koren2010collaborative} or framed them in an explore/exploit paradigm~\cite{Vanchinathan:2014,li2010contextual}.  Recommendation systems have natural connections with the explore/exploit trade-off; for example, should a system recommend items that have high probability of being consumed under the current model, or low probability items to learn more about a user's preferences?  Reinforcement learning models already use notion of a feedback loop to maximize reward.  
One major challenge with this setup, however, is constructing a reward function.  Usually the reward is based on click-through rate or revenue for companies; we, however, focus on utility for the users of a platform.  Our analysis may be informative for the construction of reward functions and may be extended to use the stationary distribution of Markov decision processes (MDPs) to further characterize the impact of the feedback loop.\looseness=-1

\section{Interaction Model}
\label{sec:interaction}
In order to reason about the feedback dynamics of algorithmic recommendations and user behavior, we need a model of how users engage with recommended items on a platform; we model engagement and not ratings, which is justified in \cref{sec:related}.
We draw on a model recently proposed by Schmit and Riquelme~\citep{schmit2017human} that captures the interaction between recommended items and users, and we modify this model to allow for personalized recommendations and multiple interactions for a given user.\footnote{Unlike the Schmit and Riquelme model \citep{schmit2017human}, we include no additional noise term because we model utility (or ``quality'' in the Schmit and Riquelme model) probabilistically instead of holding it fixed.}\looseness=-1

\begin{definition}
	The utility of user $u$ consuming item $i$ at time $t$ is
\begin{equation}
	V_{ui}(t) =  P_{ui}(t) + Q_{ui}(t),
\end{equation}
where $P_{ui}(t)$ and $Q_{ui}(t)$ are the utilities that are known and unknown to the user, respectively.
Neither utilities are known to the platform.
\end{definition}

\vspace{-3px}
When a user considers whether or not they wish to engage with items, they have some notion of their own preferences; these preferences come from any information displayed and external knowledge.  The quantity $P$ captures these preferences that are known to the user but unknown to the recommendation system platform.  Users rely on $P$ in combination with the ordering of recommended content to select items with which to engage.  Users must also have some unknown utility $Q$, or else they would be omniscient, and recommendation systems would be of no use.  The underlying objective of these systems is to estimate the total utility $V$.\looseness=-1

\begin{assumption}
The utility of a user interacting with an item is approximately static over time, or 
$V_{ui}(t) \approx V_{ui}$.
\label{as:static_utility}
\end{assumption}

\vspace{-3px}
In the real world, utility fluctuates due to contextual factors such as user mood.  However, the variance around utility is likely small and inversely related to the importance of the choice.  Moving forward, we will omit the time notation for simplicity.

\begin{assumption}
The total utility $V_{ui}$ is beta-distributed,\footnote{We use an atypical parameterization of the beta distribution with mean $\mu$ and fixed variance $\sigma$ and distinguish this parameterization as $\mbox{Beta}'$. For our simulations in \cref{sec:simulations}, we set $\sigma=10^{-5}$.  To convert to the standard parameterization, $\alpha=\left(\frac{1-\mu}{\sigma^2} - \frac{1}{\mu}\right)\mu^2$ and $\beta=\alpha\left(\frac{1}{\mu} - 1\right)$.} 
\begin{equation}
	V_{ui} \sim \mbox{Beta}'(\rho_{u} \alpha_i^\top),
\end{equation}
and is parameterized by the dot product of user general preferences $\rho_u$ for user $u$ and item attributes $\alpha_i$ for item $i$.
\label{as:utility}
\end{assumption}

\vspace{-3px}
This assumption will constrain utility values to be in the range $[0,1]$; this representation is flexible because any utility with finite support can be rescaled to fall within this range.  The use of the dot product to parameterize the utility is likewise a flexible representation; when the underlying vectors $\rho$ and $\alpha$ have a dimensionality of $\min(\vert\mathcal{U}\vert, \vert\mathcal{I}\vert)$ and either preferences $\rho$ or attributes $\alpha$ use a one-hot representation, then all possible utility values can be captured.

\begin{assumption}
General preferences $\rho$ and attributes $\alpha$ are fixed but unknown to the user or the recommendation system.  They are drawn from Dirichlet distributions, or 
\begin{equation}
\rho_u \sim \mbox{Dirichlet}(\mu_\rho)
\qquad
\mbox{and}
\qquad
\alpha_i \sim \mbox{Dirichlet}(\mu_\alpha),
\label{eq:pref}
\end{equation}
for all users $u \in \mathcal{U}$ and all items $i \in \mathcal{I}$, respectively.  Individual preferences are parameterized by a vector of global popularity of preferences $\mu_\rho$ over all users.  Individual item attributes are similarly parameterized by a global popularity of attributes $\mu_\alpha$ over all items.
\looseness=-1
\label{as:prefs}
\end{assumption}

\vspace{-3px}
\Cref{as:utility} requires $\rho_{u} \alpha_i^\top \in [0,1]$ and this guarantees that $\rho$ and $\alpha$ will satisfy this constraint.  
Most importantly, when aggregated by user (a proxy for activity) or item (popularity), this construction produces a distribution of utility values with a long tail, as seen in real platforms~\citep{celma2010long}.

\begin{assumption}
The proportion of the utility known to user $u$ is $\eta_{ui} \sim \mbox{Beta}'(\mu_{\eta})$;\footnote{Mean proportion $\mu_{\eta}=0.98$ in our simulations (\cref{sec:simulations}).}
this results in 
\begin{equation}
P_{ui} = \eta_{ui} V_{ui}
\quad
\mbox{and}
\quad
Q_{ui} = (1-\eta_{ui}) V_{ui}.
\end{equation}
\end{assumption}

\vspace{-3px}
This assumption introduces some uncertainty so that the known utility $P$ is a noisy approximation of the true utility $V$.  While each user could theoretically have a different mean proportion $\mu_{\eta}$, in practice this is not important because the known utilities $P$ are not compared across users.

\begin{assumption}
	At every discrete time step $t$, each user $u$ will interact with exactly one item, $i_{u}(t)$.
\label{as:one_item}
\end{assumption}

\vspace{-3px}
Users in the real world have varying levels of activity; 
we argue that the long tail of utility (see the justification for \cref{as:prefs}) captures the essence of this behavior and that we could adopt different levels of user activity without substantially altering our results.\looseness=-1

\begin{definition}
To select an item at time $t$, user $u$ relies on her own preferences $P_{ui}$ and a function $f$ of the rank of the items provided by recommendation system $\mathfrak{R}$.\footnote{For our simulations (\cref{sec:simulations}), we used $f(n) = n^{-0.8}$, which approximates observed effects of rank on click-through rate \cite{jansen2013effect}; our results held for other choices of $f$.}  The chosen item is
\begin{equation}
	i^\mathfrak{R}_u(t) = \arg\max_i \left( f\left(\mbox{rank}^\mathfrak{R}_{u,t}(i)\right) \cdot P_{ui}(t)  \right),
\label{eq:choice}
\end{equation}
where $\mbox{rank}^\mathfrak{R}_{u,t}(i) = n$, or according to recommender system $\mathfrak{R}$'s ordering of items, as described in \cref{sec:framework}, the $n$th recommendation $r$ for user $u$ at time $t$ is item $i$; this can also be written $r_{u,n}(t) = i$.
\end{definition}

\vspace{-3px}
Users are more likely to click on items presented earlier in a ranked list~\cite{jansen2013effect}; the function $f$ captures this effect of rank on the rate of interaction.  In keeping with our earlier discussion, to allow for various levels of user activity, one need only add some threshold $\tau$ such that if the function of rank and preference $P_{ui}(t)$ inside \cref{eq:choice} are less than this threshold, then no interaction occurs.\looseness=-1

\begin{assumption}
	Each user $u$ interacts with item $i$ at most once.
\label{as:only_once}
\end{assumption}

\vspace{-3px}
When a user engages with an item repeatedly, utility decreases with each interaction.  This is the simplest assumption that captures the notion of decreasing utility; without it, a generally poor recommendation system might ostensibly perform well due to a single item.  The interaction model could alternatively decrease utility with multiple interactions, but this would not alter results significantly.\looseness=-1

\begin{assumption}
	New and recommended items are interleaved.
\label{as:new_items}
\end{assumption}

\vspace{-3px}
As in the real world, new items are introduced with each time interval.  When no recommendation algorithm is in place (early ``start-up'' iterations), the system randomly recommends the newest items.  Once a recommendation algorithm is active, we interleave new items with recommended items; this interleaving procedure is a proxy for users engaging with items outside of the recommendation system, or elsewhere on the platform. Since this procedure is identical for all systems, it does not impact the comparison across systems.\looseness=-1

\section{Simulated Communities}
\label{sec:simulations}
In this section, we explore the performance of various recommendation systems on simulated communities of users and items.  We first describe the simulation procedure, and then discuss three claims.

\subsection{Simulation Procedure}

We consider six recommendation algorithms: \emph{popularity}, \emph{content filtering} (``content''), \emph{matrix factorization} (``MF''), \emph{social filtering} (``social''), \emph{random}, and \emph{ideal}.
\Cref{sec:framework} provides further details on the first four approaches and describes our general recommendation framework.  The core idea of this framework is that each recommendation system provides some underlying score $s_{ui}(t)$ of how much a user $u$ will enjoy item $i$ at time $t$.  These scores are constructed using user preferences $\theta_u(t)$ and item attributes $\beta_i(t)$:
\begin{equation}
s_{ui}(t) = \theta_u(t) \beta_i(t)^\top,
\label{eq:rec_score_main}
\end{equation} and each recommendation approach has a different way of constructing or modeling these preferences and attributes.\looseness=-1

For our simulations, all of the six approaches recommend from the set of items that exist in the system at the time of training; \emph{random} recommends these items in random order.  \emph{Ideal} recommends items for each user $u$ based on the user's true utility $V_{ui}$ for those items.  Comparison with these two approaches minimizes the impact of the interaction model assumptions (\cref{sec:interaction}) on our results.\looseness=-1

\begin{figure}[tb]
\includegraphics[width=\columnwidth]{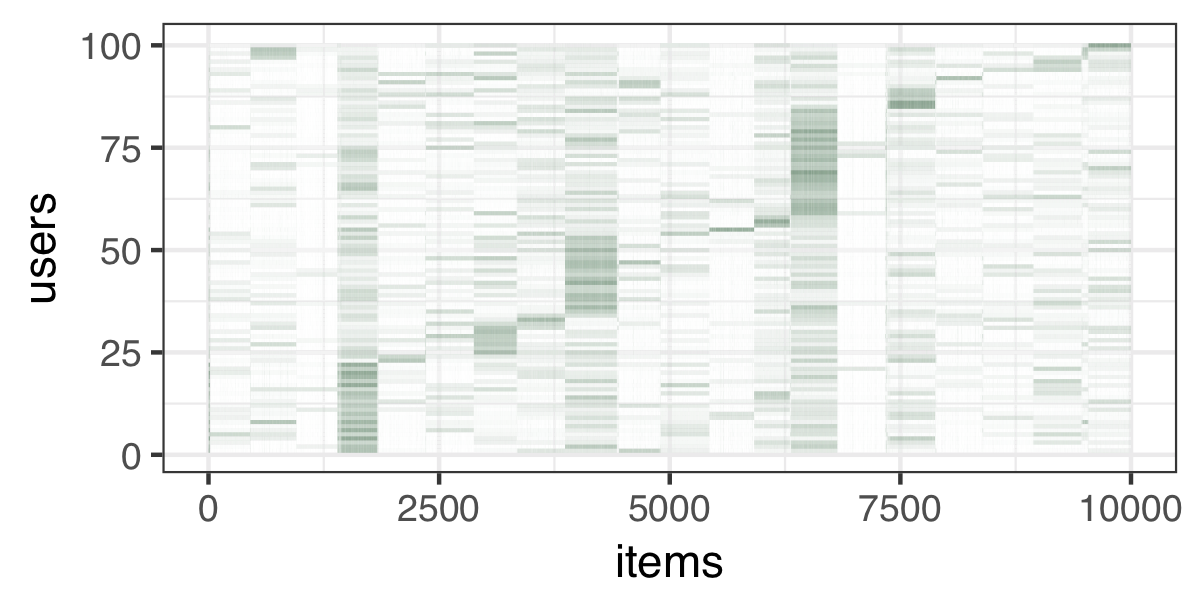}
\caption{Example true utility matrix $V$ for simulated data; darker is higher utility.  The distribution of user preferences is disproportionate, like the real world, and the structure is easily captured with matrix factorization.}
\label{fig:world}
\end{figure}

In all of our simulations, a community consists of $100$ users and is run for 1,000 time intervals with ten new items being introduced at each interval; each simulation is repeated with ten random seeds and all our results are averages over these ten ``worlds.''
We generate the distributions of user preference and item attribute popularity, as used in \cref{eq:pref}, in $K=20$ dimensions; we generate uneven user preferences, but approximately even item attributes.  The user preference parameter is generated as follows:
$\tilde\mu_\rho \sim \mbox{Dirichlet}(\boldsymbol{1})$ and $\mu_\rho = 10 \cdot \tilde\mu_\rho$.
This mirrors the real world where preferences are unevenly distributed, which allows us to expose properties of the recommendation algorithms.  Item attribute popularity is encouraged to be more even in aggregate, but still be sparse for individual items; we draw: $\tilde\mu_\alpha \sim \mbox{Dirichlet}(\boldsymbol{100})$ and $\mu_\alpha = 0.1 \cdot \tilde\mu_\alpha$.
While item attributes are not evenly distributed in the real world, this ensures that all users will be able to find items that match their preferences.  With these settings for user preferences and item attributes, the resulting matrix of true utility is sparse (e.g., \cref{fig:world}), which matches commonly accepted intuitions about user behavior.
We generate social networks (used by social filtering only) using the covariance matrix of user preferences; we impose that each user must have at least one network connection and binarize the covariance matrix using this criteria.  This procedure enforces homophily between connected users, which is generally (but not always) true in the real world.\looseness=-1

We consider two cases of observing user interactions with items: a simple case where each recommendation algorithm is trained once, and a more complicated case of repeated training; this allows us to compare a single cycle of the feedback loop (\cref{fig:cartoon}) to multiple cycles.  In the simple paradigm, we run 50 iterations of ``start-up'' (new items only each iteration), train the algorithms, and then observe 50 iterations of confounded behavior.  In the second paradigm, we have ten iterations of ``start-up,'' then train the algorithms every iteration for the remaining 90 iterations using all previous data.

\subsection{Homogenization Effects}
\label{sec:homogenization}

\begin{figure*}[tb]
\includegraphics[width=\textwidth]{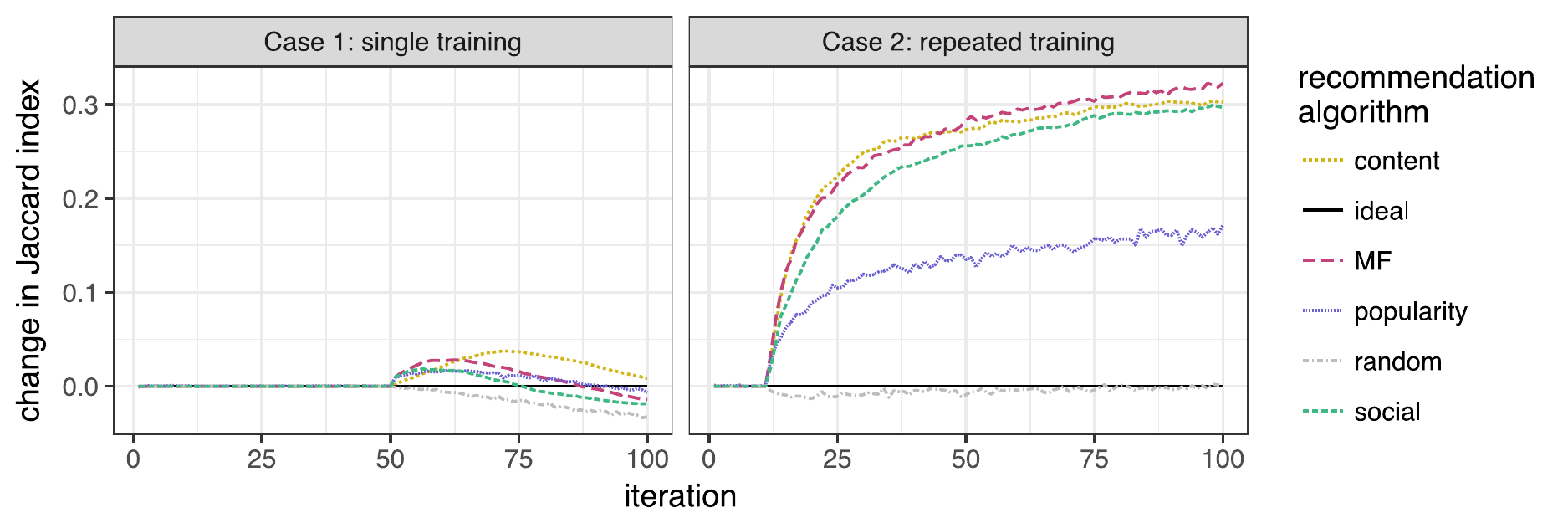}
\caption{Change in Jaccard index of user behavior relative to ideal behavior; users paired by cosine similarity of $\theta$.  On the left, mild homogenization of behavior occurs soon after a single training, but then diminishes.  On the right, recommendation systems that include repeated training homogenize user behavior more than is needed for ideal utility.}
\label{fig:jaccard_time}
\end{figure*}
 
\begin{figure}[tb]
\includegraphics[width=\columnwidth]{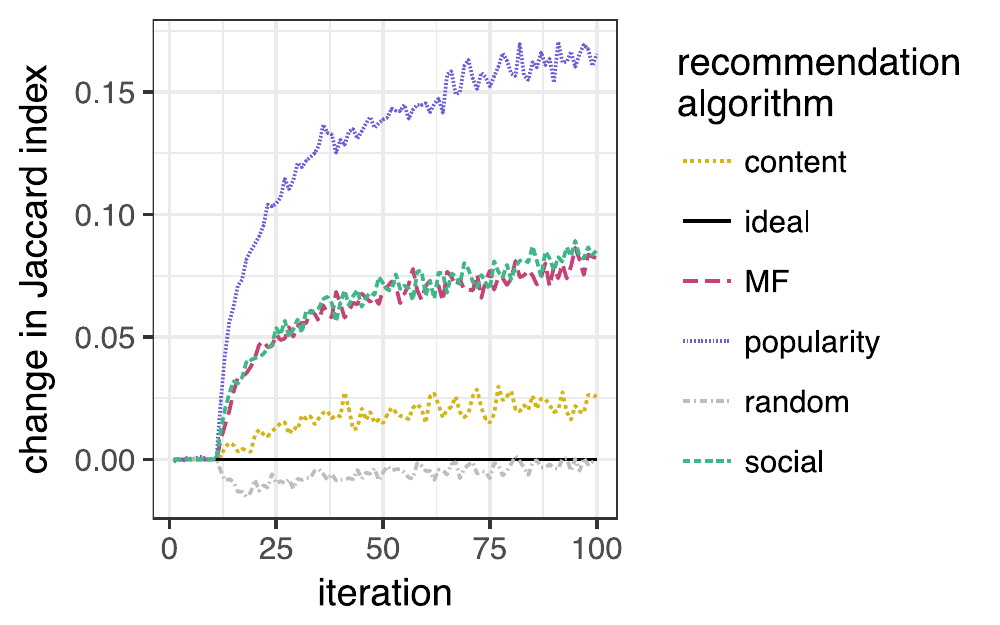}
\caption{For the repeated training case, change in Jaccard index of user behavior relative to ideal behavior; users paired randomly.  Popularity increases homogenization the most globally, but all non-random recommendation algorithms also homogenize users globally.}
\label{fig:jaccard_global}
\end{figure}

\begin{figure*}[t]
\includegraphics[width=\textwidth]{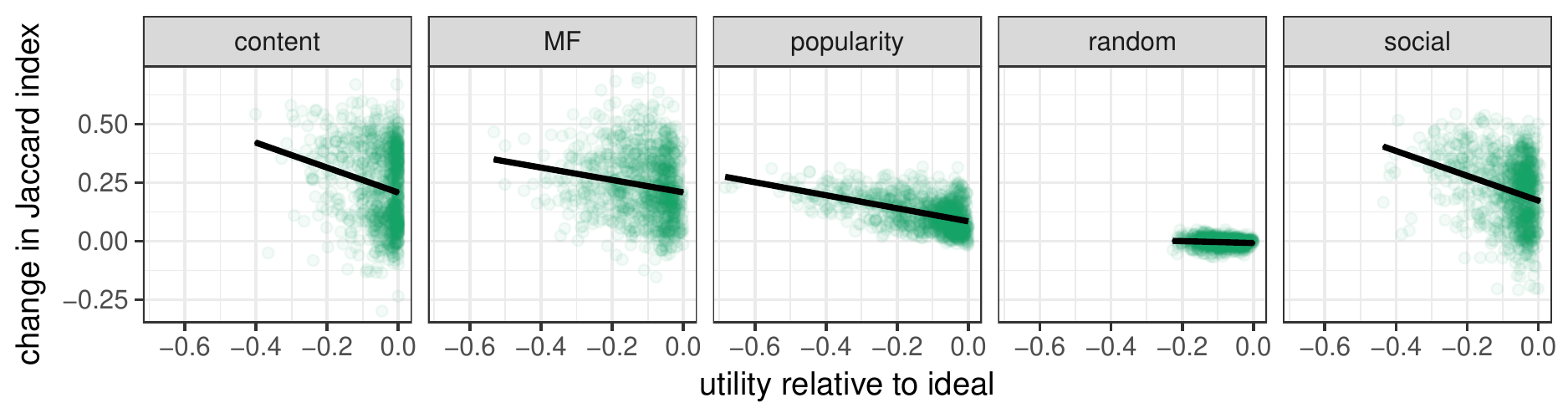}
\caption{For the repeated training case, change in Jaccard index of user behavior, relative to ideal behavior, and shown as a function of utility relative to the ideal platform; users paired by cosine similarity of $\theta$. Each user is shown as a point, with a linear fit to highlight the general trend that users who experience losses in utility have higher homogenization.}
\label{fig:jaccard_util}
\end{figure*}

Recommendation systems may not change the underlying preferences of user (especially not when used on short time scales), but they do impact user behavior, or the collection of items with which users interact.  Recommendation algorithms encourage similar users to interact with the same set of items, therefore homogenizing their behavior, relative to the same platform without recommended content. For example, \emph{Popularity-based} systems represent all users in the same way; this homogenizes all users, as seen in previous work~\cite{celma2008hits,treviranus2009value}.  \emph{Social} recommendation systems homogenize connected users or within cliques, and \emph{matrix factorization} homogenizes users along learned latent factors.

Homogenizing effects are not inherently bad as they indicate that the models are learning patterns from the data, as intended; when achieving optimum utility, users will have some degree of homogenization.  However, homogenization of user behavior does not correspond directly with an increase in utility: we can observe an increase in homogenization without a corresponding increase in utility.  This is related to the explore/exploit paradigm, where we wish to exploit the user representation to maximize utility, but not to homogenize users more than necessary.  When a representation of users is over-exploited, users are being pushed to be have more similar behaviors than their underlying preferences would optimally dictate.  This suggests that the ``tyranny of majority'' and niche ``echo chamber'' effects may both be manifestations of the same problem: over-exploitation of recommendation models.  While concerns about maximizing utility are not new to the recommendation system literature, there are also grave social consequences from homogenization that have received less consideration.\looseness=-1

We measure homogenization of behavior by first pairing each user with their most similar user according to the recommendation system, or user $u$ is partnered with user $v$ that maximizes the cosine similarity of $\theta_u$ and $\theta_v$.  Next, we compute the Jaccard index of the sets of observed items for these users.  If at time $t$, user $u$ has interacted with a set if items $\mathcal{D}_u(t) = \{i_u(1), \dots, i_u(t)\}$ the Jaccard index of the two users' interactions can be written as
\begin{equation}
\mathbf{J}_{uv}(t) = \frac{\vert \mathcal{D}_u(t) \cap \mathcal{D}_v(t) \vert}{\vert \mathcal{D}_u(t) \cup \mathcal{D}_v(t) \vert}.
\end{equation}
We compared the Jaccard index for paired users against the Jaccard index of the same users exposed to ideal recommendations; this difference captures how much the behavior has homogenized relative to ideal.  \Cref{fig:jaccard_time} shows these results for both the single training and the repeated training cases.
In the single training case, users became slightly homogenized after training, but then returned to the ideal homogenization.  With repeated training, all recommendation systems (except random), homogenize user behavior beyond what was needed to achieve ideal utility.  As the number of cycles in the feedback loop (\cref{fig:cartoon}) increases, we observe homogenization effects continue to increase without corresponding increases in utility.\looseness=-1

We can consider global homogenization to reveal the impact of the feedback loop at the population level; instead of comparing to paired users based on $\theta_u$, we compare users matched randomly (\cref{fig:jaccard_global}).  In this setting, all recommendation systems (except random) increased global homogeneity of user behavior.  The popularity system increased homogeneity the most; after that, matrix factorization and social filtering homogenized users comparably, and content filtering homogenized users least of all, but still more than ideal.\looseness=-1

We have shown that when practitioners update their models without considering the feedback loop of recommendation and interaction, they encourage users to consume a more narrow range of items, both in terms of local niche behavior and global behavior.\looseness=-1

\subsection{Loss of Utility}
\label{sec:utility}

Changes in utility due to these effects are not necessarily born equally across all users.  For example, users whose true preferences are not captured well by the low dimensional representation of user preferences may be disproportionately impacted.  These minority users may see lesser improvements or even decreases in utility when homogenization occurs.  \Cref{fig:jaccard_util} breaks down the relationship between homogenization and utility by user; for all recommendation algorithms, we find that users who experience lower utility generally have higher homogenization with their nearest neighbor.

Note that we have assumed that each user/item pair has fixed utility (\cref{as:static_utility}).  In reality, a collection of similar items is probably less useful than a collection of diverse items \cite{drosou2017diversity}.  With a more nuanced representation of utility that includes the collection of interactions as a whole, these effects would likely increase in magnitude.\looseness=-1

\subsection{Shifts in Item Consumption}
\label{sec:gini}
To explore the impacts of these systems on item consumption, we computed the Gini coefficient on the distribution of consumed items, which measures the inequality of the distribution (0 being all items are consumed at equal rates and 1 indicating maximal inequality between the consumption rate of items).  We use the following definition of Gini inequality, which relies on the relative popularity (RP)\footnote{RP$(i,t)$ is the rank of item $i$ based on how many people have consumed it, relative to all other items (up to time $t$).} of each item $i$ at time $t$:
\begin{equation}
    \mathcal{G}(t) = \frac{\sum_{i\in \mathcal{I}}\left(2 * \mbox{RP}(i,t) - \vert\mathcal{I}\vert - 1\right)\sum_{u\in \mathcal{U}}\mathds{1}\left[i \in \mathcal{D}_u(t)\right]}{\vert\mathcal{I}\vert\sum_{i\in \mathcal{I}}\sum_{u\in \mathcal{U}}\mathds{1}\left[i \in \mathcal{D}_u(t)\right]}.
\end{equation}

\begin{figure}[t]
\includegraphics[width=\columnwidth]{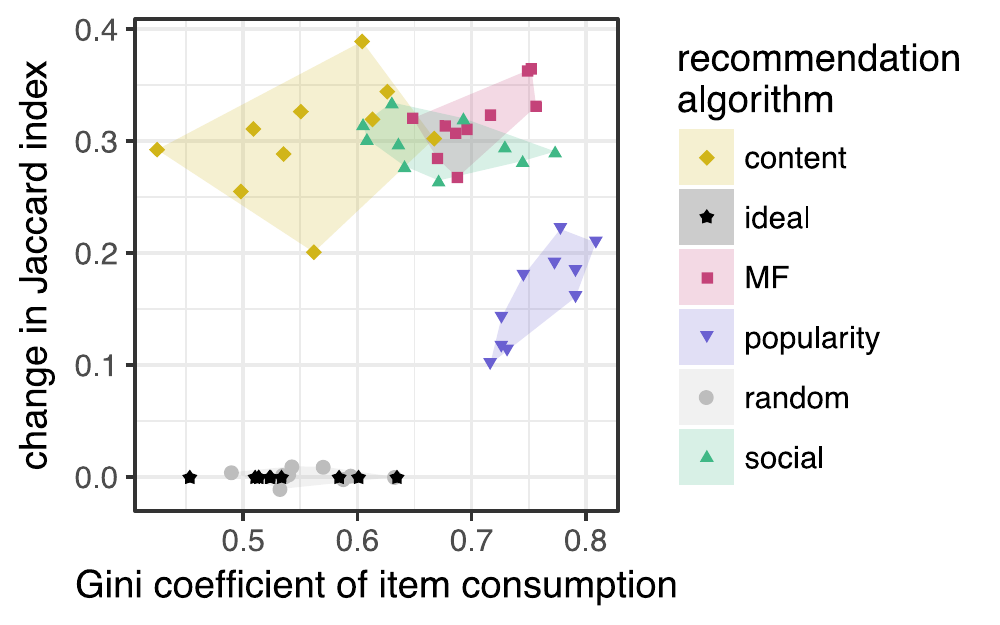}
\caption{For the repeated training case, change in Jaccard index of user behavior (higher is more homogeneous), relative to the Gini coefficient of the item consumption distribution (higher is more unequal consumption of items). Each point is a single simulation.  Similar homogenization can result in different item consumption distributions.}
\label{fig:gini}
\end{figure}

We found that the feedback loop amplifies the impact of recommendation systems on the distribution of item consumption, irrespective of homogenization effects.  Specifically, two recommendation systems can produce similar amounts of user homogenization with different distributions of item consumption (\cref{fig:gini}).  For example, matrix factorization (MF) and content filtering have comparable homogenizing effects, but MF creates a more unequal distribution of item consumption.

As a community, we do not fully understand the ways in which these systems change the popularity of items.
Differential item consumption may ultimately change item production from strategic actors, such as companies like Amazon and Netflix which are now producing content based on their consumers' behavior data.  Thus, recommendation systems change not only what users see first, but can fundamentally alter the collection of content from which users can choose.\looseness=-1

\section{Accounting for Confounding}
\label{sec:confounded_eval}
Recommendation system are often evaluated using confounded held-out data.  This form of offline evaluation is used by both researchers and practitioners alike, but the choice is not as innocuous as one might expect.  
When a recommendation system is evaluated using confounded held-out data, results are biased toward recommendation systems similar to the confounding algorithm.
Researchers can, intentionally or unintentionally, select data sets that highlight their proposed model, thus overstating its performance.

There are specific instances in recommendation system literature where this may be troubling.  For example, \emph{MovieLens} is a research website that uses collaborative filtering to make movie recommendations \citep{miller2003movielens}; the researchers behind it have released several public datasets\footnote{\url{https://grouplens.org/datasets/movielens/}} that are used prolifically in evaluation recommendation systems \citep{harper2016movielens}.  Recommendation systems based on collaborative filtering will likely perform better on this data.
Similarly, the \emph{Douban} service contains popularity features; algorithms that include a notion of popularity should perform better on the publicly available data from this platform~\cite{hao:sr}.
Many algorithms use social network information to improve recommendations~\cite{Chaney:2015,guo2015trustsvd,hao:sr} but when training and evaluating these algorithms on data from social platforms, it is not clear if the models capture the true preferences of users or the biasing effects of platform features.
These biases are problematic for researchers and platform developers who attempt to evaluate models offline or rely on academic publications that do.\looseness=-1

Researchers and practitioners alike would benefit from methods to address these concerns.  Weighting techniques, such as those proposed by 
Schnabel, et al.~\cite{schnabel2016recommendations} and De Myttenaere, et al.~\cite{de2014reducing} for offline evaluation seem promising.  We performed a preliminary exploration of weighting techniques and found that in the repeated training case, weighting can simultaneously increase utility and decrease homogenization.  These weighting techniques could also be of use when attempting to answer social science questions using algorithmically confounded user behavior data.  We leave a full exploration of these methods for future work. 

As proposed by Bottou, et al.~\cite{bottou2013counterfactual}, formal causal inference techniques can assist in the design of deployed learning systems to avoid confounding.  This would likely reduce the effects we have described (\cref{sec:simulations}), but needs to be studied in greater depth.  In particular, the distribution of impact across users should be explored, as well as non-accuracy metrics. Regardless, practitioners would do well to incorporate measures to avoid confounding, such as these.  At least they should log information about the recommendation system in deployment along with observations of behavior; this would be useful in disentangling recommendation system influence from true preference signal as weighting and other techniques are refined.\looseness=-1

\section{Conclusion}
\label{sec:conclusion}
We have explored the impact of algorithmic confounding on a range of simulated recommendation systems. We found that algorithmic confounding amplifies the homogenization of user behavior (\cref{sec:homogenization}) without corresponding gains in utility (\cref{sec:utility}) and also amplifies the impact of recommendation systems on the distribution of item consumption (\cref{sec:gini}).
These findings have implications for any live recommendation platform; those who design these systems need to consider how a system influences its users and account for algorithmic confounding. Researchers who use confounded data to test and evaluate their algorithms should be aware of these effects, as should researchers who wish to use confounded data to study and make claims about user behavior.  Platform users and policy makers should take these effects into consideration as they make individual choices or propose policies to guide or govern the use of these algorithms.

\section*{Acknowledgements} This work was funded by a Princeton CITP Interdisciplinary Microsoft Seed Grant. BEE was funded by NIH R01 HL133218, a Sloan Faculty Fellowship, and an NSF CAREER AWD1005627.

\appendix
\section{Recommendation Framework}
\label{sec:framework}
In this appendix, we cast ostensibly disparate recommendation methods into a general mathematical framework that encompasses many standard algorithmic recommendation techniques.

A recommendation system provides some underlying score $s_{ui}(t)$ of how much a user $u$ will enjoy item $i$ at time $t$.  This score is generated from two components: the system's representations at time $t$ of both user preferences $\theta_u(t)$ for user $u$ and item attributes $\beta_i(t)$ for item $i$.  The dot product of these vectors produces the score:
\begin{equation}
s_{ui}(t) = \theta_u(t) \beta_i(t)^\top.
\label{eq:rec_score}
\end{equation}
This construction is reminiscent of the notation typically used for the matrix factorization approach discussed in \cref{sec:collaborative_filtering}, but it also provides us a more general framework.

The scores $s_{ui}(t)$ are not necessarily comparable across recommendation techniques.  Recommendation systems that focus on \emph{rating prediction} will provide scores comparable to other rating prediction models.  For these systems, the goal is to predict how users will explicitly rate items, most commonly on a five-star scale, and are typically evaluated with prediction error on held-out ratings.
Low errors for predicting ratings, however, do not always correspond to high accuracy in rank-based evaluation \citep{cremonesi2010performance,steck2013evaluation} (e.g., ``what should I watch next?''), which is the ultimate goal in many recommendation applications.  Additionally, limiting our scope to rating prediction systems would omit models that focus on learning rankings \citep{karatzoglou2013learning} or that otherwise produce rankings directly, such as ordering items by popularity.

Given a collection of scores $s_{ui}(t)$, a recommendation system then produces, for each user, an ordered list (or sequence) of items sorted according to these scores. Formally, we represent these recommendations as a set of sequences
\begin{equation}
\left\{ \big( r_{u,n}(t) \big)_{n=1}^{|\mathcal{I}|} \right\}_{u\in \mathcal{U}},
\end{equation}
where $\mathcal{I}$ is the set of all items and $\mathcal{U}$ is the set of all users.  For each user $u$, the system provides a sequence, or ranked list of items, where $n$ is the position in the ranked list and $r_{u,n}$ is the recommended item for user $u$ at rank $n$.  This sequence of items for a given user $u$ is defined as all items sorted descendingly by their respective score $s_{ui}(t)$, or
\begin{equation}
\big( r_{u,n}(t) \big)_{n=1}^{|\mathcal{I}|} = \mbox{descending sort} \big( \forall i \in \mathcal{I}, \mbox{by}=s_{ui}(t)\big).
\label{eq:rank}
\end{equation}
Our simulated experiments (\cref{sec:simulations}) revealed that it is important to break ties randomly when performing this sort; if not, the random item recommender baseline receives a performance advantage on early iterations by exposing users to a wider variety of items.

We now cast a collection of standard recommendation systems in this framework by defining the user preferences $\theta_u(t)$ and item attributes $\beta_i(t)$ for each system; this emphasizes the commonalities between the various approaches.

\subsection{Popularity}
\label{sec:pop}
Intuitively, the popularity recommendation system ranks items based on the overall item consumption patterns of a community of users.
All users are represented identically, or $\theta_{u}(t) = 1$ for all $u \in \mathcal{U}$, and thus every user receives the same recommendations at a given time $t$.
Item attributes are based on the interactions of users with items up to time $t$; these interactions can be structured as a collection of $N$ triplets $\left\{ (u_n,i_n,t_n) \right\}_{n=1}^N$, where each triplet $(u,i,t)$ indicates that user $u$ interacted with item $i$ at time $t$.

There are many permutations of the popularity recommendation technique, including windowing or decaying interactions to prioritize recent behavior; this prevents recommendations from stagnating. For our analysis (\cref{sec:simulations}), we employ the simplest popularity recommendation system; it considers all interactions up to time $t$, or\looseness=-1
\begin{equation}
\beta_{i}(t) = \sum_{n=1}^{N} \mathds{1}\left[i_n = i \mbox{ \sc{and} } t_n < t \right].\footnote{The indicator function notation $\mathds{1}[\mbox{ \sc{expression} }]$ evaluates to 1 when the internal expression is true and 0 when it is false.}
\label{eq:pop_simple}
\end{equation}

\subsection{Content Filtering}
\label{sec:content}
Content-based recommender systems match attributes in a user's profile with attribute tags associated with an item \citep[ch. 3]{RSH}.  In the simplest variant, the set of possible attribute tags $\mathcal{A}$ are identical for both users and items.
User preferences $\theta_u(t)$ is a vector of length $|\mathcal{A}|$, and the attribute tags for a given user are in the set $\mathcal{A}_u(t)$; this gives us
$
\theta_{ua}(t) = \mathds{1}\left[a \in \mathcal{A}_u(t) \right]$.
The item attribute tags can similarly be represented as a vector of length $|\mathcal{A}|$ with values
$
\beta_{ia}(t) = \mathds{1}\left[a \in \mathcal{A}_i(t) \right]$,
where $\mathcal{A}_i(t)$ is the set of attributes for an item $i$.\looseness=-1

Attributes for both users and items can be input manually (e.g., movie genre), or they can be learned independent of time $t$ with, for example, a topic model \citep{Blei:2012} for text or from the acoustic structure of a song \citep{tingle2010exploring} for music; when learned, the attribute tags can be real-valued instead of binary.
For our simulations (\cref{sec:simulations}), we use binary item attributes and learn real-valued user preferences.\footnote{Item attributes are determined by applying a binarizing threshold to $\alpha_i$ in \cref{eq:pref} such that every item has at least one attribute.  User representations are then learned using \texttt{scipy.optimize.nnls} \citep{scipy}.}

\subsection{Social Filtering}
\label{sec:social}
Social filtering recommendation systems rely on a user's social network to determine what content to suggest.  In the simplest variant, the user preferences $\boldsymbol{\theta}(t)$ are a representation of the social network, or a $|\mathcal{U}| \times |\mathcal{U}|$ matrix; for each user $u$ connected to another user $v$, $\theta_{uv}(t) = \mathds{1}\left[v \in \mathcal{F}_u(t) \right]$,
where $\mathcal{F}_u(t)$ is the set of people $u$ follows (directed network) or with which they are connected as ``friends'' (undirected network) as of time $t$. Alternatively, the user preferences $\boldsymbol{\theta}(t)$ can represent the non-binary trust between users, which can be provided explicitly by the user \citep{Massa2007} or learned from user behavior \citep{Chaney:2015}; we use the latter in our analysis (\cref{sec:simulations}).

Item attributes $\boldsymbol{\beta}(t)$ are then a representation of previous interactions, broken down by user, or an $|\mathcal{I}| \times |\mathcal{U}|$ matrix where for each item $i$ and user $v$,
$\beta_{iv}(t) = \mathds{1}\left[v \in \mathcal{U}_i(t) \right]$,
where $\mathcal{U}_i(t)$ is the set of users which have interacted with item $i$ as of time $t$.
The item representation $\boldsymbol{\beta}(t)$ can alternatively be a non-binary matrix, where $\beta_{iv}(t)$ is the number of interactions a user $v$ has with an item $i$, or user $v$'s rating of item $i$.

\subsection{Collaborative Filtering}
\label{sec:collaborative_filtering}
Collaborative filtering learns the representation of both users and items based on past user behavior, and is divided into roughly two areas: neighborhood methods and latent factor models.  

\paragraph{Neighborhood Methods}
The simplicity of neighborhood methods \citep[ch. 4]{RSH} is appealing for both implementation and interpretation.  These approaches find similar users, or neighbors, in preference space; alternatively, they can find neighborhoods based on item similarity.  In either case, these methods construct similarity measures between users or items and recommend content based on these measures.  We outline the user-based neighborhood paradigm, but the item-based approach has a parallel construction.\looseness=-1

Users are represented in terms of their similarity to others, or
\begin{equation}
\theta_{u}(t) = \left[w_{u1}, \dots, w_{u \vert\mathcal{U}\vert}\right],
\end{equation}
where the weight $w_{uv}$ captures the similarity between users $u$ and $v$.  The similarity between users is typically computed using their ratings or interactions with items, and there are many options for similarity measures, including Pearson's correlation, cosine similarity, and Spearman's rank correlation \cite{ahn2008new}.  These weights can be normalized or limited to the closest $K$ nearest neighbors.

Items are represented with their previous interactions or ratings, just as done for social filtering in \cref{sec:social}.  We can see that these neighborhood methods are very similar to social filtering methods---the biggest distinction is that in social filtering, the users themselves determine the pool of users that contribute to the recommendation system, whereas the collaborative filtering approach determines this collection of users based on similarity of behavior.

While neighborhood-based methods have some advantages, we focus our analysis of collaborative filtering approaches on latent factor methods for two main reasons: first, in the simulated setting, there is little distinction between social filtering and neighborhood-based collaborative filtering.  Second, latent factor methods are more frequently used than neighborhood methods.

\paragraph{Latent Factor Methods.}
Of the latent factor methods, matrix factorization is a successful and popular approach \citep{Koren09}. 
The core idea behind matrix factorization for recommendation is that user-item interactions form a $|\mathcal{U}| \times |\mathcal{I}|$ matrix (as of time $t$) $\mathbf{R}(t)$, which can be factorized into two low-rank matrices: a $|\mathcal{U}| \times K$ representation of user preferences $\boldsymbol{\theta}(t)$ and an $|\mathcal{I}| \times K$ representation of item attributes $\boldsymbol{\beta}(t)$.  The number of latent features $K$ is usually chosen by the analyst or determined with a nonparametric model.  The multiplication of these two low-rank matrices approximates the observed interaction matrix, parallel to \cref{eq:rec_score}, or
$\boldsymbol{\theta}(t) \boldsymbol{\beta}(t)^\top\approx\mathbf{R}(t)$.

There are many instantiations of the user preferences and item attributes.
Non-negative matrix factorization \citep{Lee00} requires that these representations be non-negative.  Probabilistic matrix factorization \citep{PMF} assumes that each cell in the user preference and item attribute matrices are normally distributed, whereas a probabilistic construction of non-negative matrix factorization \citep{CannyGaP} assumes that they are gamma-distributed.  Under all of these constructions, these latent representations are learned from the data by following a sequence of updates to infer the parameters that best match the observed data.\looseness=-1

Other latent factor methods, such as principal component analysis (PCA) \citep{jolliffe2002principal} and latent Dirichlet allocation (LDA) \citep{Blei:2003}, similarly frame user and item representations in terms of a low dimension $K$ of hidden factors.  These factors are then learned algorithmically from the observed interactions. We focus on matrix factorization in our simulations (\cref{sec:simulations}), for simplicity.\footnote{Specifically, we use Gaussian probabilistic matrix factorization with confidence weighting, as described by Wang and Blei~\citep{CTR}, with $a=1$ and $b=0.001$.}

To update a latent factor recommendation system with recently collected data, one has several options.  Since these methods are typically presented without regard to time $t$, one option is to refit the model entirely from scratch, concatenating old data with the new; in these cases, consistent interpretation of latent factors may be challenging.
A second option is to hold fixed some latent factors (e.g., all item attributes $\boldsymbol{\beta}$) and update the remaining factors according to the update rules used to originally learn all the latent factors.  This approach maintains the ordering of the latent factors, but may break convergence guarantees, even if it works well in practice.
This approach does not explicitly address new items or users, often called the ``cold-start problem,'' but can be adapted to account for them.\looseness=-1

\subsection{Modifications and Hybrid Approaches}
\label{sec:hybrid}

The concepts of these systems can be modified and combined to create innumerable permutations.  In considering collaborative filtering alone, neighborhood-based methods can be merged with latent factor approaches \cite{Koren:2008}.  Full collaborative filtering system can be supplemented with content information \cite{CTR} or augmented with social filtering \cite{Chaney:2015}.  Any of these methods can be supplemented with a popularity bias.  Under any of these modified or hybrid systems, the changes  propagate to the representations of user preferences $\boldsymbol{\theta}(t)$ and item attributes $\boldsymbol{\beta}(t)$, and the general framework for recommendation remains the same.

\bibliographystyle{acm}
\bibliography{library.bib}

\end{document}